\documentclass[12pt,a4paper]{article}

\usepackage[T1]{fontenc}
\usepackage[utf8]{inputenc}
\usepackage[english]{babel}
\usepackage[margin=1in]{geometry}
\usepackage{amsmath,amssymb}
\usepackage{graphicx}
\usepackage{textcomp}
\usepackage{url}
\usepackage[autostyle,english=american]{csquotes}
\MakeOuterQuote{"}
\title{Inversion symmetry breaking by interlayer hybridization in a tetralayer van der Waals heterostructure}

\author{\parbox{0.95\textwidth}{\centering\normalsize
Juan C. Arias-Muñoz*,
Cuiju Yu,
Nianze Shang,
Christopher P. Edwards,
Zixuan Ning,
Farzaneh Bouzari,
Rodolfo Canet-Albiach,
Seyed Hossein Hosseini-Shokouh,
Ruihuan Fang,
Hanlin Fang,
Sanshui Xiao,
Jose L. Lado,
Zhipei Sun*
}}
\date{}

\begin{document}

\maketitle

\begin{flushleft}
\small
J. C. Arias-Muñoz, Dr. N. Shang, C. P. Edwards, Z. Ning, F. Bouzari, Dr. R. Canet-Albiach, Dr. S. H. Hosseini-Shokouh, Dr. R. Fang, Prof. Z. Sun\\
Department of Electronics and Nanoengineering, Aalto University, Espoo 02150, Finland\\
Email Address: zhipei.sun@aalto.fi, juancamilo.arias@aalto.fi

Dr. C. Yu, Prof. J. L. Lado \\
Department of Applied Physics, Aalto University, Espoo 02150, Finland\\

Dr. H. Fang, Prof. S. Xiao \\
Department of Electrical and Photonics Engineering, Technical University of Denmark, Kongens Lyngby 2800, Denmark
\end{flushleft}

\begin{abstract}

Inversion symmetry breaking unlocks a broad range of physical phenomena fundamental for modern technologies, motivating efforts to engineer noncentrosymmetric materials. In this context, monolayer transition metal dichalcogenides, which naturally lack inversion symmetry, have emerged as a versatile platform for valley physics, strong excitonic effects, and nonlinear optics. However, in their hexagonal stacking phase, inversion symmetry is restored in even-layer stacks and bulk form, suppressing these effects and limiting applications. In this work, inversion symmetry breaking is demonstrated in a vertically stacked tetralayer heterostructure assembled from two intrinsically centrosymmetric bilayer flakes, as evidenced by second harmonic generation measurements with an intensity comparable to that of their constituent monolayer. Spectroscopic measurements and theoretical estimations indicate that effective symmetry breaking in the tetralayer heterostructure is driven by interlayer hybridization, giving rise to pronounced electronic and phononic signatures. This symmetry breaking further induces polar anisotropy, with an intensity ratio reaching 2.36. We propose that asymmetric contributions from excitons to the second harmonic optical dipole may influence this behavior. These findings establish interlayer coupling as a critical mechanism for introducing inversion symmetry breaking and tuning the optical response of van der Waals heterostructures, providing a powerful strategy for photonics and optoelectronics in two-dimensional materials.

\end{abstract}


\section{Introduction}

Breaking crystal inversion symmetry is one of the fundamental pillars of a wide range of modern technologies, such as piezoelectrics \cite{zhu_observation_2015, wu_piezoelectricity_2014}, electro-optic modulators \cite{sun_optical_2016}, ferroelectrics \cite{yasuda_stacking_2021, fei_ferroelectric_2018}, and chiral optics \cite{ zhang_coherent_2022, mak_valley_2012, cao_valley-selective_2012}. Although intrinsic noncentrosymmetric materials are widely used, the demand for advanced integrable platforms has increasingly driven interest toward artificial inversion symmetry breaking as an effective strategy for engineering next-generation functional materials and devices \cite{du_engineering_2021,vizner_interfacial_2021,wang_differential_2026}. Within this framework, monolayer (1L) two-dimensional (2D) van der Waals (vdW) materials have recently emerged as a promising building block for engineering the properties of integrated optoelectronic devices and exploring symmetry–quantum correlations beyond the limitations of conventional bulk materials \cite{novoselov_2d_2016,turunen_quantum_2022}. Particularly in transition metal dichalcogenides (TMDs), valley polarization brought about by their noncentrosymmetric lattice structure and strong spin-orbit coupling allows us to study the interplay of space and time reversal symmetry. This enables us to harness their unique electronic and photonic responses, such as pronounced nonlinear optical (NLO) effects \cite{autere_nonlinear_2018,hsu_second_2014}. However, the most commonly occurring hexagonal polytype (2H) of TMDs does not exhibit inversion symmetry breaking in even-numbered layers or at the bulk limit without considering complex phase information \cite{tang_forbidden_2026}. This configuration suppresses many symmetry-dependent physical properties, such as even-order NLO responses \cite{li_probing_2013,arias-munoz_interlayer_2024} and valley-contrasting phenomena, thereby limiting their broader application in nonlinear photonic devices.

To access the underlying exotic physical phenomena in 2D centrosymmetric materials, it is necessary to introduce inversion symmetry breaking. In fact, inducing inversion symmetry breaking in centrosymmetric 2D materials through stacking is conceptually straightforward, but thus far has been primarily demonstrated in graphene-based structures \cite{cao_unconventional_2018,yang_tunable_2020,zhang_emergent_2023}. However, a stacked TMD system can exhibit effective inversion symmetry breaking only when strong interlayer interactions are present, ideally sufficient to perturb the intrinsic coupling  within each constituent layer \cite{arias-munoz_interlayer_2024} and thereby break the structural symmetry.

In this work, we experimentally demonstrate inversion symmetry breaking in vertically stacked centrosymmetric 2H-TMDs for the first time. Second harmonic generation (SHG) measurements in a tetralayer MoS$_2$/WSe$_2$ heterostructure provide direct evidence of the broken inversion symmetry. We attribute the symmetry breaking to interlayer hybridization observed as strong mechanical and excitonic coupling, which gives rise to new collective interlayer vibrational modes and excitonic resonances. The strong coupling provides new properties distinct from those of the individual centrosymmetric bilayer (2L) building blocks, as evidenced by Raman spectroscopy. Photoluminescence (PL) spectroscopy further reveals strong interlayer excitonic coupling arising from interlayer interactions. The SHG measurements also reveal anisotropic NLO polarization associated with the broken inversion symmetry, with an anisotropy ratio of approximately 2.36, compared to the isotropic behavior in 1L samples. An interlayer coupling dependent tetralayer system potentially could offer a 100\% modulation depth of the SHG signal at new excitonic energies originating from interactions between the two constituent materials. Compared with heterobilayers, this platform is more robust against perturbations and disorder, while potentially offering a new degree of tunability and improved control over emergent optical and nonlinear properties. Additionally, the emerging anisotropy in the nature of the SHG signal introduces a new degree of freedom for light modulation and potential optical switching applications. Our results highlight interlayer coupling engineering  as a promising mechanism for breaking inversion symmetry in otherwise centrosymmetric TMDs, enabling the targeted tuning of physical properties through interlayer interactions.

\section{Results}

\subsection{Evidence of inversion symmetry breaking in a tetralayer vdW heterostructure}\label{subsec1}

\begin{figure}[htbp]
    \centering
    \includegraphics[width=\linewidth]{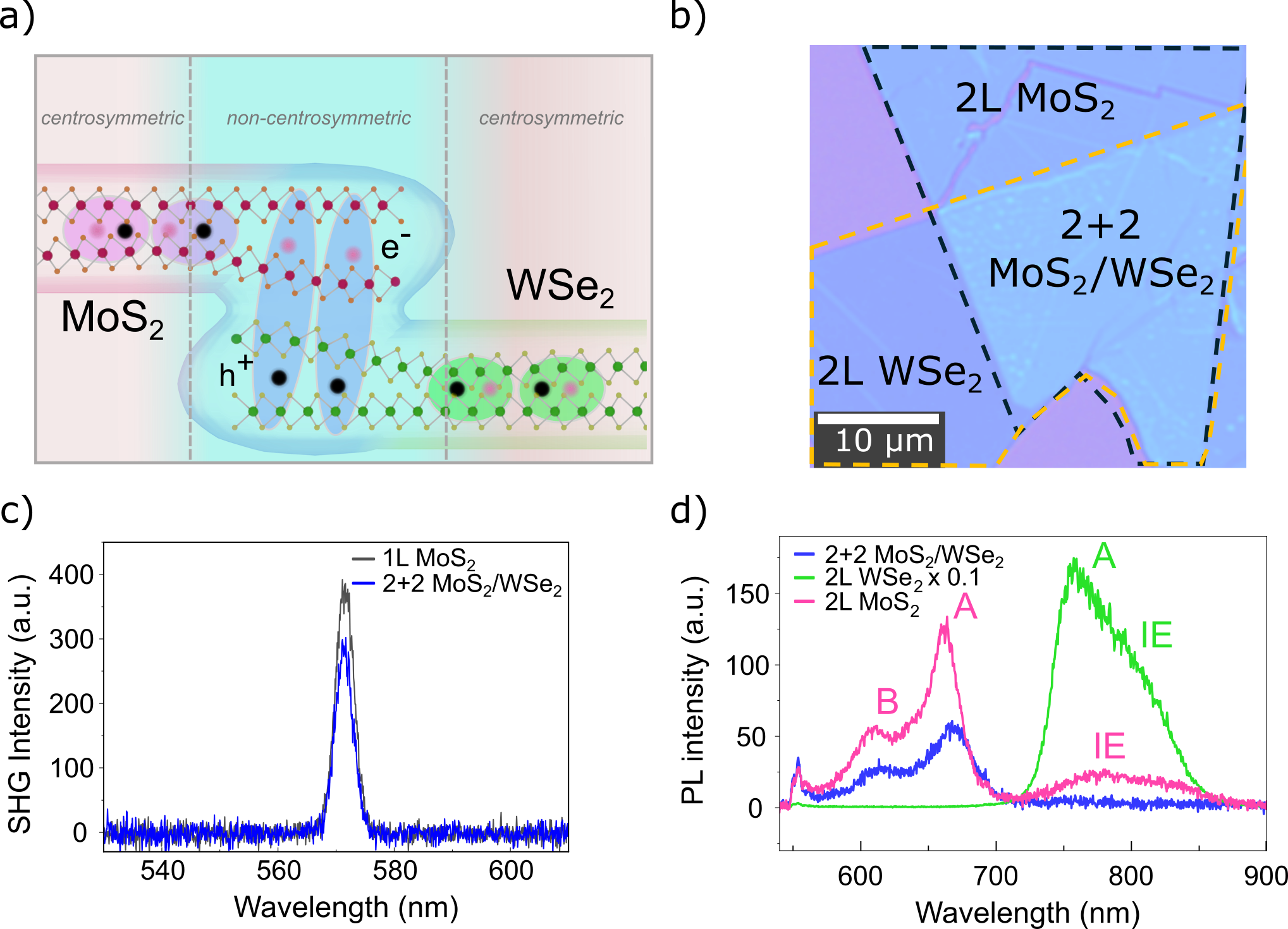}
    \caption{\textbf{Inversion symmetry breaking in a tetralayer heterostructure.} \textbf{a)} Side view representation of the vertical heterostructure indicating the centrosymmetric and noncentrosymmetric regions. Intralayer excitons are depicted in pink and green for MoS$_2$ and WSe$_2$, respectively. The IX is indicated in blue. \textbf{b)} Optical microscopy image of the heterostructure. The three studied areas (i.e., 2L WSe$_2$, 2L MoS$_2$, and 2+2 MoS$_2$/WSe$_2$ ) are labeled and the individual flakes are highlighted with dashed lines for clarity. \textbf{c)} Inversion symmetry breaking in a tetralayer heterostructure, evidenced by SHG (under 1150~nm incident light). The signal from a 1L MoS$_2$ is presented for comparison. \textbf{d)} Intralayer PL quenching in the visible range from the tetralayer heterostructure in b). PL from intrinsic bilayer WSe$_2$ and MoS$_2$ are shown for comparison with identified direct (A-, B-) and indirect (IE) excitons. The signal from 2L WSe$_2$ is reduced by a factor of 10 for representation purposes.}
    \label{fig_IX}
\end{figure}

In stacked TMDs, inversion symmetry breaking is typically accompanied by strong excitonic coupling and the emergence of interlayer excitons (IX) \cite{du_engineering_2021} due to strong interlayer interactions and the resulting charge redistribution between layers. To detect IX signatures from the heterostructure, the PL from both intralayer excitons in the constituent flakes and the IX must be simultaneously resolvable. To this end, we employ a type-II band alignment, in which electrons and holes relax in different materials, rather than homostructures or heterostructures of similar band gaps. A schematic representation of the IX formation in a tetralayer vdW heterostructure and the inversion symmetry breaking induced by the proposed stacking configuration is depicted in \textbf{Figure} \ref{fig_IX}a.

Our heterostructure consists of two heterogeneous centrosymmetric building blocks: a 2L 2H-MoS$_2$ flake and a 2L 2H-WSe$_2$ flake (\textbf{Figure} \ref{fig_IX}a and b). The heterostructure is obtained by mechanical exfoliation and vertical dry transfer techniques, with an angle between flakes of either $\sim$3.808±1.02° or $\sim$56.192±1.02°. As discussed later, stacks near 0° and 60° exhibit similar room-temperature electronic structures. This, combined with the absence of intrinsic SHG in 2L TMDs, prevents a direct distinction between 0° and 60° configurations in the tetralayer. To maximize the coupling between flakes, we maintained a clean interface between TMD layers (Supplementary Figure 1). Air pockets visible in Figure \ref{fig_IX}b are attributed to slight variations in the transfer speed.

Direct evidence of inversion symmetry breaking in the system is provided by comparing the SHG response of the heterostructure with that of 1L MoS$_2$ (\textbf{Figure} \ref{fig_IX}c). Under the same excitation conditions, the heterostructure shows an SHG signal comparable to the 1L, confirming effective inversion symmetry breaking in the stacked region. A detailed analysis of the nature of the heterostructure SHG signal is presented later in this work and in the Supplementary Information.

The evolution of the intralayer excitonic recombination after stacking is investigated by PL measurements (\textbf{Figure} \ref{fig_IX}d) carried out in three areas corresponding to the two intrinsic 2L TMDs and the heterostructure. In Figure \ref{fig_IX}d the PL signal associated with the direct recombination of A-excitons, appears at approximately 660~nm for MoS$_2$ (pink curve) and 759~nm for WSe$_2$ (green curve) \cite{arora_excitonic_2015,splendiani_emerging_2010}. In the case of MoS$_2$, the less intense signal at 610~nm from B-excitons can also be observed as a result of the valence band splitting due to spin-orbit coupling \cite{kosmider_large_2013}. The intrinsic 2L TMD areas also present PL signal from indirect recombination (IE) at 786~nm and 793~nm for MoS$_2$ and WSe$_2$, respectively, related to indirect $\Lambda-\Gamma$ recombination pathways \cite{zhao_origin_2013,kunstmann_momentum-space_2018}.

In the tetralayer area (i.e., 2+2 MoS$_2$/ WSe$_2$ area, blue curve), the signal corresponding to the PL of the intrinsic 2L flakes is significantly diminished. The PL contribution from the MoS$_2$ A- and B-excitons redshifts $\sim$10~nm and $\sim$5~nm, respectively, while its IE signature at 786~nm becomes significantly suppressed. Notably, the PL signal in the visible range from WSe$_2$ is also suppressed to a level below the detection capability of our instruments. The strong PL quenching, as compared to individual 2L MoS$_2$ and WSe$_2$ reflects strong interlayer coupling, which is also typically observed in the case of 2L heterostructures \cite{chiu_determination_2015}.

\begin{figure}[htbp]
    \centering
    \includegraphics[width=\linewidth]{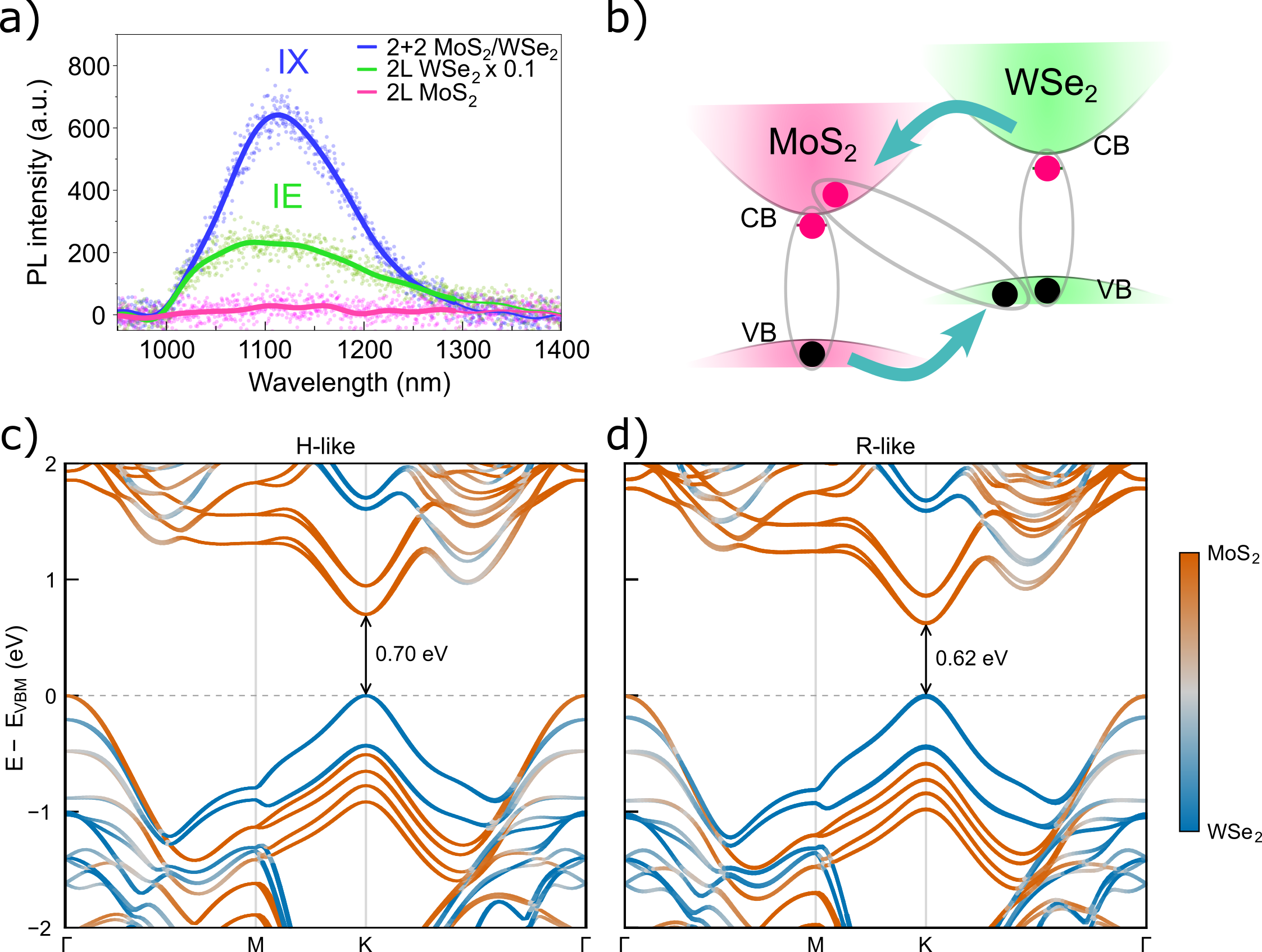}
    \caption{\textbf{Observation of strong interlayer coupling in a tetralayer vdW heterostructure.} \textbf{a)} Appearance of the IX PL signal from the tetralayer heterostructure. PL from intrinsic 2L WSe$_2$ and MoS$_2$ are shown for comparison. The IE exciton from WSe$_2$ and the IX exciton from the tetralayer heterostructure are labeled. \textbf{b)} Simplified representation of the IX formation, showing the CBM and VBM bands from the stacked materials.  \textbf{c)} Band structure estimations from the heterostructure for H-like (60°) stacking configuration and \textbf{d)} for R-like (0°) stacking configuration. The bands are colored by the weight on the MoS$_2$ and WSe$_2$ layers.}
    \label{fig_DFT}
\end{figure}

We also observe a strong infrared (IR) PL signal from the tetralayer heterostructure at 1114~nm (\textbf{Figure} \ref{fig_DFT}a), which we attribute to an interlayer-excitonic contribution. IR PL signal of the same energy has also been reported in a MoS$_2$/WSe$_2$ 2L heterostructure corresponding to the $K-K$ recombination pathway \cite{lin_moire-engineered_2024}, however, in our structure, this distinction is not clear (Supplementary Note 6). Additionally, WSe$_2$ manifests a second IE peak at 1090~nm, corresponding to the indirect $K-\Gamma$ transition, arising from the near-degeneracy of the conduction band minima at the $\Lambda$ and $K$ points. To further confirm that the signal from the heterostructure originates from the IX, we measure a 2L reference heterostructure formed by 1L MoS$_2$ and 1L WSe$_2$ (PL details in Supplementary Figure 2). Density Functional Theory (DFT) calculations further reveal the interlayer character of the band gap from the tetralayer heterostructure, for stackings at 0° and 60° (Supplementary Figure 3). The IX PL wavelengths from the 2L and tetralayer heterostructures are in good agreement with the expectations for twist angles of $\sim3.808^\circ\pm1.02^\circ$ (or $\sim56.192^\circ\pm1.02^\circ$) \cite{tan_layer-engineered_2021,lin_moire-engineered_2024}, i.e., beyond the regime where the moiré potential becomes significant.

The appearance of the IX PL signal (Figure \ref{fig_DFT}a), along with the quenching of the PL signal in the visible region of the spectrum (Figure \ref{fig_IX}d), indicate strong interlayer interaction and introduces an energy transfer process that favors interlayer recombination over intralayer recombination processes \cite{kunstmann_momentum-space_2018}. A schematic of the carrier dynamics between flakes is depicted in \textbf{Figure} \ref{fig_DFT}b.

Band structure analysis reveals the electronic dispersion in the heterostructure, which gives rise to an interlayer band gap between the valence band maximum (VBM) of WSe$_2$ and the conduction band minimum (CBM) of MoS$_2$ (\textbf{Figure} \ref{fig_DFT}c and d). The calculated interlayer K--K gap (0.70~eV, PBE+SOC) lies below the measured IR emission at 1.11~eV, as expected because PBE underestimates quasiparticle gaps. The calculation uses independent particles, without electron--hole interactions or local-field effects, and therefore contains no excitons. Within the uncertainty of the calculated interlayer alignment (up to 0.3~eV for the strained commensurate cell), the calculations confirm the type-II alignment and that the IR PL arises from interlayer recombination between the flakes. They cannot establish which interlayer transition is involved, since the K and $\Gamma$ valence maxima are degenerate within our accuracy (Supplementary Note 6).

\subsection{Interlayer coupling probed by Raman spectroscopy}\label{subsec2}

\begin{figure}[htbp]
    \centering
    \includegraphics[width=\linewidth]{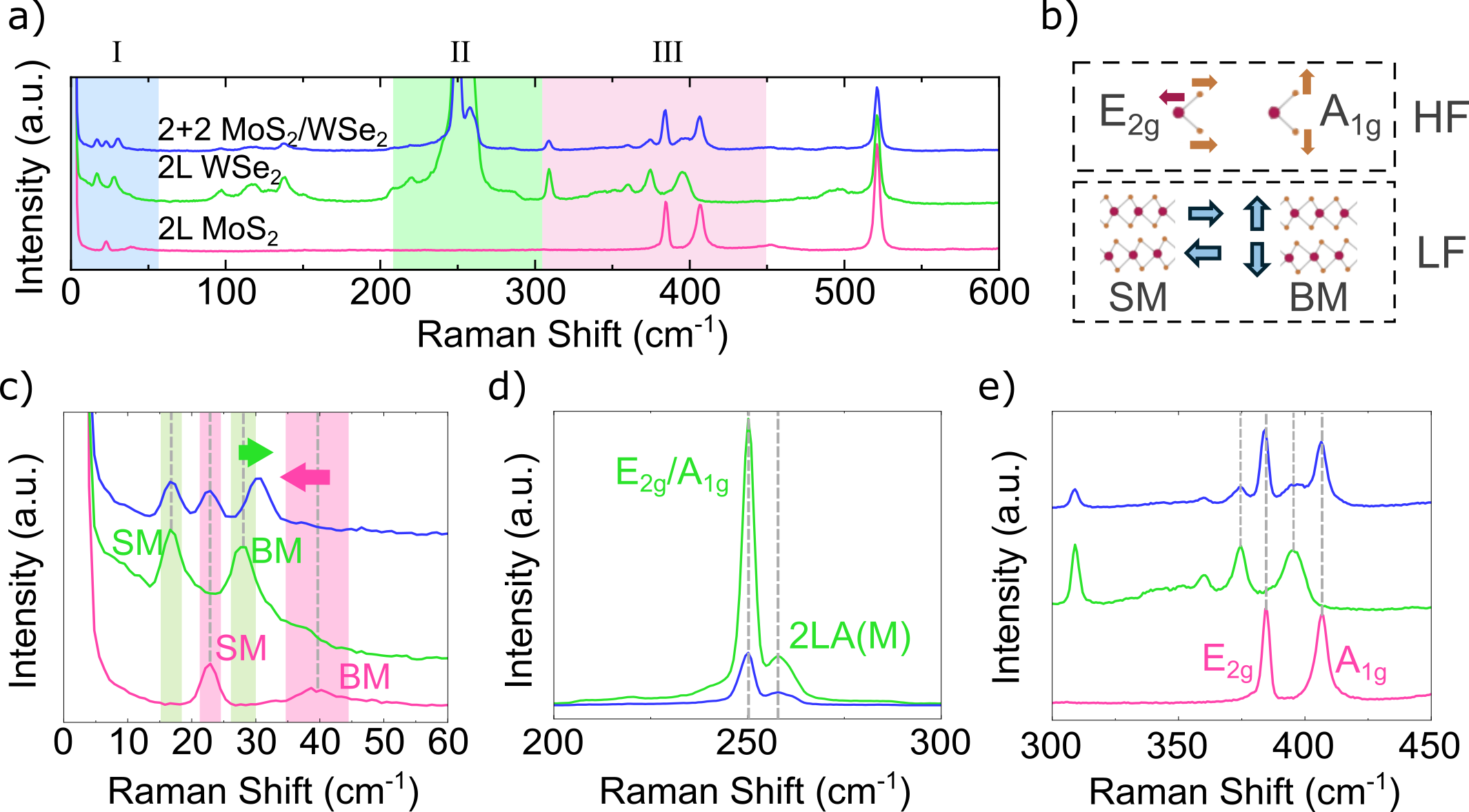}
    \caption{\textbf{Evolution of vibrational modes due to interlayer coupling.} \textbf{a)} Raman spectra from the three regions of interest. The spectra presented in c), d) and e) correspond to the highlighted regions I, II and III, respectively. \textbf{b)} Schematic representation of the HF (E$_{2g}$ and A$_{1g}$) and LF (BM and SM) vibrational modes. \textbf{c)} Magnified results of the LF mode region (Region I) in a). Arrows show the relative frequency shifts from the intrinsic BM positions to the heterostructure BM position. \textbf{d)} Overlay of the HF modes from WSe$_2$ and the heterostructure in Region II in a). \textbf{e)} Magnified results of Region III in a) highlighting the HF modes from MoS$_2$.}
    \label{fig_Raman}
\end{figure}

To further understand the mechanism of inversion symmetry breaking, we investigate the vibrational modes in the intrinsic flakes and their transformation in the heterostructure. \textbf{Figure} \ref{fig_Raman}a presents the Raman spectra from the three regions, emphasizing the high-frequency (HF) modes from the component flakes in Regions II (WSe$_2$) and III (MoS$_2$), and the low-frequency (LF) modes in Region I. A schematic representation of the vibrational modes in TMDs is provided in \textbf{Figure} \ref{fig_Raman}b, depicting the oscillations of the LF (lower panel) shearing (SM) and breathing (BM) modes, as well as the HF (upper panel) E$_{2g}$ and A$_{1g}$ modes.

We study the relative oscillations of the atomic layers in the heterostructure reflected in the LF Raman modes \cite{yan_stacking-dependent_2015,arias-munoz_interlayer_2024,obrien_mapping_2016}. \textbf{Figure} \ref{fig_Raman}c shows that the SM of 2L WSe$_2$ (16.62±0.13~cm$^{-1}$, green curve) and MoS$_2$ (22.89±0.01~cm$^{-1}$, pink curve) are inherited by the heterostructure (blue curve) and are present with relatively similar intensities. In contrast, the BM from the heterostructure shows a new oscillation frequency in the stacked region. This new frequency is localized at 30.5±0.08~cm$^{-1}$, intermediate between the positions of the 2L MoS$_2$ BM mode centered at 39.65±0.25~cm$^{-1}$, and the 2L WSe$_2$ BM mode centered at 27.95±0.09~cm$^{-1}$. Remarkably, this interlayer BM at 30.5±0.08~cm$^{-1}$ has been previously observed in MoS$_2$/WSe$_2$ 2L heterostructures \cite{kim_anomalous_2022}. The presence of the unique BM indicates strong interlayer coupling between the contacting faces of the stacked layers. The estimated out-of-plane coupling strength is $\sim$74\% of the intrinsic interlayer coupling strength of the constituent 2L (Supplementary Note 2). This results in reorganized collective out-of-plane layer oscillations, while the layers within different materials retain nearly intrinsic in-plane motion.

Higher-energy Raman modes reveal the atomic motions within individual flakes. Changes in intensity ratios in these HF modes are related to structural modifications in the system \cite{velicky_strain_2020,kim_anomalous_2022}. To observe the evolution of the modes, we analyze HF Regions II and III in Figure \ref{fig_Raman}a. In Region II (200~-~300~cm$^{-1}$, \textbf{Figure} \ref{fig_Raman}d), a reduction in the intensity of the~WSe$_2$ HF A$_{1g}$ and E$_{2g}$ peaks is observed, although due to their proximity in energy they are not resolvable from each other. The decrease in signal intensity is consistent with the Raman signatures of thicker WSe$_2$ flakes under 532~nm excitation \cite{blaga_unveiling_2024,Tonndorf_photoluminescence_13}, indicating strong interaction between materials. Voigt curve fittings (Supplementary Note~2 and Supplementary Figure 4) reveal a redshift of 0.35±0.05~cm$^{-1}$ for WSe$_2$ HF modes and of 0.6±0.02~cm$^{-1}$ and 0.38±0.06~cm$^{-1}$ for the MoS$_2$ E$_{2g}$ and A$_{1g}$ modes, respectively. The shift of the E$_{2g}$ in-plane modes ( Regions II and III, \textbf{Figure} \ref{fig_Raman}e), although minimal, suggests strain  in the range of $\sim0.1\text{--}0.21\%$ at the interface \cite{li_efficient_2020}(Supplementary Note 2). Given its in-plane nature, this mode is more sensitive to changes in lattice geometry than the A$_{1g}$ mode \cite{doan_charge_2017,velicky_strain_2020}. Nevertheless, the simultaneous redshift of the A$_{1g}$ mode indicates that the strain experienced by the flakes is tensile \cite{yagmurcukardes_strain_2018,li_efficient_2020}, as expected from the lattice mismatch between flakes. The reduction in intensity of the Raman signal from MoS$_2$ is also observed in 2L MoS$_2$/WSe$_2$ heterostructures at angles close to 60° \cite{kim_anomalous_2022}, confirming strong interlayer coupling in the tetralayer.

The mode evolution indicates a strong out-of-plane phonon interaction in the heterostructure, dominated by WSe$_2$, while in-plane interlayer and atomic oscillations within individual TMD layers remain nearly intrinsic. These observations suggest that at a twist angle of $3.808$±$1.02$° the system presents strong interlayer coupling while trending to a close-to-incommensurate configuration due to the lattice mismatch and weakened moir\'e potential.

\subsection{Detailed study of inversion symmetry breaking by SHG spectroscopy}\label{subsec3}

In this section, we examine the SHG response of the tetralayer heterostructure in detail as SHG probing is extremely sensitive to electric dipole formation in noncentrosymmetric media. \textbf{Figure} \ref{fig_SHG}a shows the SHG response of the tetralayer heterostructure, confirmed by power-dependent SHG in Supplementary Figure 6. The SHG intensity from the heterostructure ($I_{SHG}$) is comparable to that of a reference 1L MoS$_2$ on the same substrate, as depicted in Figure \ref{fig_IX}c, but is at least 10 times larger than the signal from the 2L areas (Figure \ref{fig_SHG}a). Additionally, we map the SHG response across the region of interest (\textbf{Figure} \ref{fig_SHG}b) localizing the origin of the signal to the stacked region with several "dark" spots corresponding to areas of weak coupling.  The effect of interlayer coupling strength on SHG signal is estimated in the Supplementary Note 5. It is worth noting that despite being intrinsically centrosymmetric, the 2L MoS$_2$ and WSe$_2$ regions exhibit extremely faint SHG (also shown in \textbf{Figure} \ref{fig_SHG}a and c), which is still detectable in our experiments. Such a weak SHG signal has been previously observed \cite{zhao_atomically_2016,shree_interlayer_2021}, and is typically attributed to the dielectric asymmetry between the top and bottom layers of the flakes.

\begin{figure}[htbp]
    \centering
    \includegraphics[width=0.75 \linewidth]{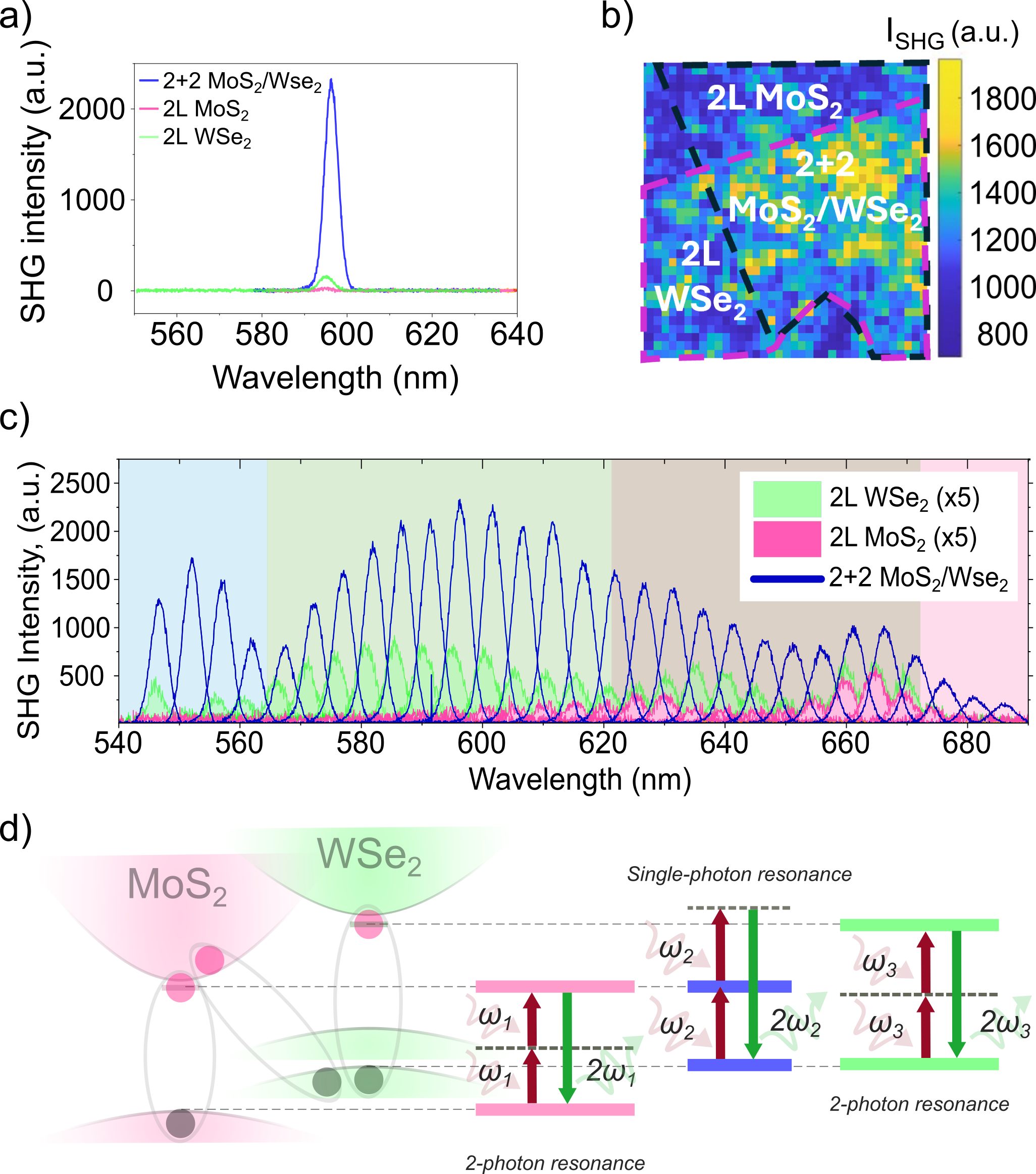}
    \caption{\textbf{Observation of inversion symmetry breaking in the tetralayer heterostructure by SHG.} \textbf{a)} SHG from the tetralayer heterostructure region under 1190 nm incident light. The faint SHG signals from 2L flakes are also provided for comparison. \textbf{b)} SHG scan from the studied area under 1200~nm excitation. Different flakes are delimited by dashed lines as a guide for the eye. \textbf{c)} Wavelength-dependent SHG response of the tetralayer heterostructure. The faint signals from the individual centrosymmetric 2L flakes are provided for comparison and are multiplied by a factor of five for clarity. The exciton-resonant contributions from MoS$_2$ (pink region), WSe$_2$ (green region) and the IX (blue region) are shaded in the colors corresponding to their respective optical transition processes depicted in d). \textbf{d)} Schematic representation of the exciton-enhanced SHG process in the heterojunction for $\hbar\omega_2>\hbar\omega_3>\hbar\omega_1$. Single-photon IX resonance, with photon energy $\hbar\omega_2$, is depicted in blue. Double-resonant processes are depicted in pink and green (with photon energies $\hbar\omega_1$ and $\hbar\omega_3$, respectively). The electronic band diagram on the left acts as a guide for interpreting real (solid lines) and virtual (dark dashed lines) states in the optical processes.}
    \label{fig_SHG}
\end{figure}

By sweeping the energy of the incident photons, we observe a wavelength-dependent SHG response from the tetralayer heterostructure, depicted in Figure \ref{fig_SHG}c, with the SHG intensity enhanced by $\sim$10-30~times across the excitation range. In the heterostructure spectrum (blue curve), the MoS$_2$ dominated segment (pink Region) and the WSe$_2$ dominated segment (green Region) exhibit contributions from their corresponding intralayer excitons. SHG signal maxima are observed at 585~nm for the B-exciton of WSe$_2$ (green curve) and at 625~nm and 665~nm for the B- and A-excitons of MoS$_2$ (pink curve), respectively. WSe$_2$ also presents two additional local maxima at 635~nm and 665~nm corresponding to its 3\textit{s} and 2\textit{s} exciton states, respectively \cite{aslan_strain_2018,chen_luminescent_2019}. The features inherited by the heterostructure show broadening and redshifting of the excitonic resonances, reaching their maximum SHG intensity at an emission wavelength of 595~nm. We attribute these phenomena to the fast charge transfer between flakes, and the renormalization of electronic states at the heterojunction \cite{paradisanos_second_2022}. Moreover, a distinct local maximum unique to the heterostructure is evident in the 540–565~nm range (blue Region), which we attribute to a single-photon resonance with the IX occurring when the incident photon energy falls within the measured IX energy range (Figure \ref{fig_DFT}a).

In the single-photon resonance scenario, an incident photon excites an electron-hole pair, separating the hole in the valence band of WSe$_2$ and the electron in the conduction band of MoS$_2$, while the second photon excites the electron further into a higher energetic virtual state before recombination. This contrasts with the double-resonant process, where two incident photons are required to excite the carriers to the conduction band. The described mechanisms are schematically illustrated in \textbf{Figure} \ref{fig_SHG}d.

To estimate the second order susceptibility ($\chi^{(2)}$), which quantifies the polarizability of a material, we implement the surface SHG theory described in \cite{shen_optical_1989} and applied in SI units in \cite{woodward_characterization_2016,autere_optical_2018} (more details in Supplementary Note~3). \textbf{Figure} \ref{fig_pSHG}a compares our results with DFT calculations of the effective $|\chi^{(2)}_{yyy}|$ element without and with excitonic effects (Bethe--Salpeter equation, BSE). Including excitons moves the calculated maximum to 2.03~eV (H-like) and 2.08~eV (R-like), close to the maximum of the measured $|\chi^{(2)}_{\mathrm{eff}}|$ at 2.05~eV, while the calculated magnitude remains larger than measured (Supplementary Note 6). Our results confirm different SHG responses at intralayer and interlayer excitonic regimes with a susceptibility of the order of 1L MoS$_2$. Further details on the theoretical model and the $\chi^{(2)}$ estimations are provided in the Methods section and in Supplementary Note~3, respectively.

To characterize the in-plane tensor component and the polarizability of the heterostructure, we measure its polarization-dependent SHG (pSHG) under 1100 nm incident light. The results reveal a characteristic 6-fold symmetric component (expected from the minimal misalignment between flakes) overlapped with a notably stronger 2-fold component along the 0° ($-30$°) direction in parallel (perpendicular) configuration (\textbf{Figure} \ref{fig_pSHG}b). The anisotropic feature presents a rotation of 19.2 ± 2.1° if measured under 1200~nm (Supplementary Figure 8). The relatively low strain revealed by Raman measurements, and the different wavelength-dependent angular responses exclude strain as the unique source of this anisotropy. We attribute the observed behavior to the coherent interference from the symmetry and anisotropy components of the SHG in the spectral region resonant with the interlayer exciton (Supplementary Note 4).

To quantify the anisotropy introduced, we deconstruct the effective susceptibility tensor in symmetry and anisotropy contributions and obtained the coherent SHG response for the parallel and perpendicular radiating channels (Supplementary Note 4).  For each polarization channel we define the anisotropic signal enhancement as the ratio between the major and minor radii of the pSHG envelope $\rho_{\parallel}=\frac{r^{maj}_\parallel}{r^{min}_\parallel} \approx \frac{r^{maj}_\perp}{r^{min}_\perp}= \rho_{\perp}$. The SHG intensity in each radiating channel is given by:

\begin{equation}\nonumber
I_{\parallel,\perp}(\theta)=I_{0}
\left|\mathcal{A}^{\mathrm{symm}}_{\parallel,\perp}(\theta)
+\mathcal{A}^{{anis}}_{{\parallel,\perp}}(\theta)\right|^2
\end{equation}

The relative phase between projected amplitudes determines
the interference and the deformation added to the six-fold pattern of the pSHG signal.

The polar plots in Figure \ref{fig_pSHG}b present an anisotropy ratio of 2.36 formed by the interference of the  2-fold and 6-fold amplitudes.  We propose a possible origin to these observations from asymmetric contributions to the optical dipole from excitons from different locations of the Brillouin zone, since the energy difference between recombination pathways is minimal (Supplementary Note 6). However a detailed understanding of the nature of this phenomenon requires a further study that considers the effects of the twist angle and layer number on the SHG response, which lies beyond the scope of the present work.

\begin{figure}[htbp]
    \centering
    \includegraphics[width=\linewidth]{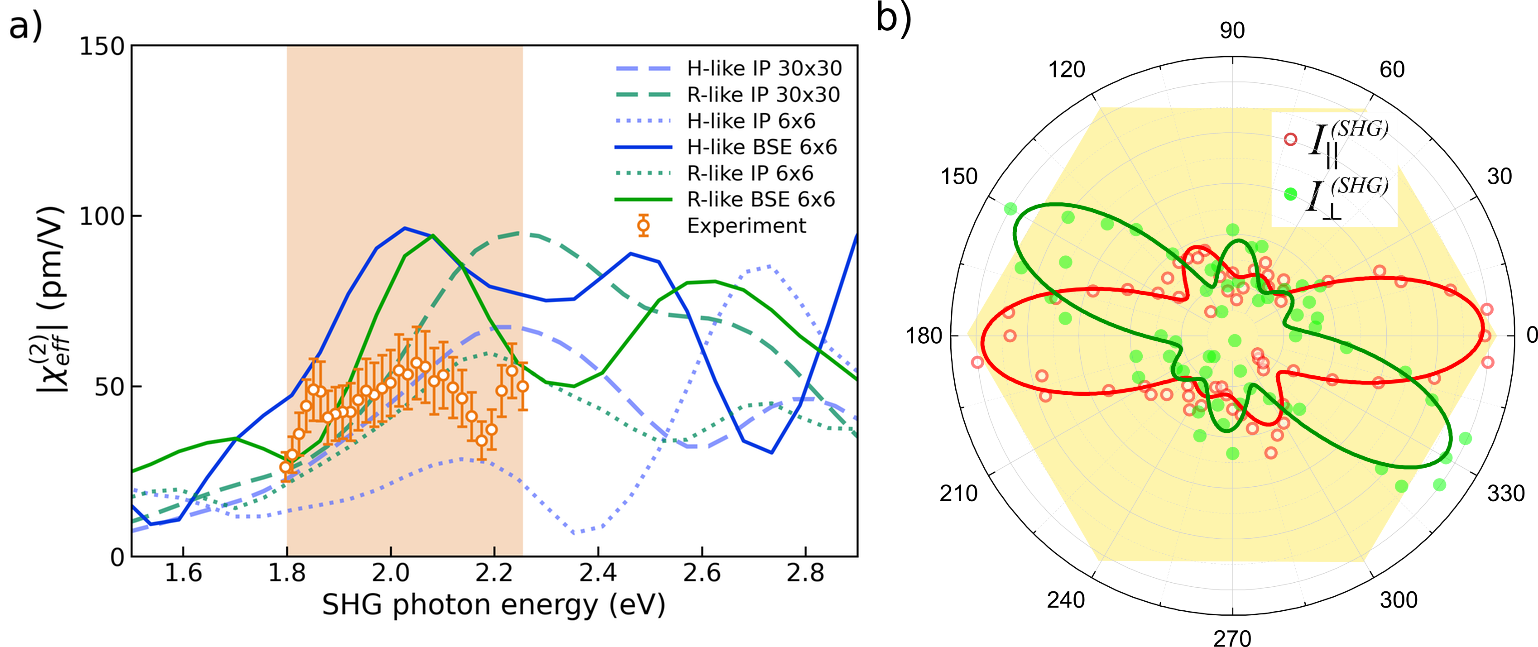}
    \caption{\textbf{Second-order NLO excitonic signature of the tetralayer heterostructure.} \textbf{a)}~Experimentally obtained effective susceptibility $|\chi^{(2)}_{eff}|$ (orange) and calculated effective thickness-normalized susceptibility, for ideal 60° (H-like) and 0° (R-like) stackings, included as reference. The orange area represents our experimentally detectable range. Solid (dashed and dotted) lines indicate calculations including (omitting) excitonic effects; dashed (dotted) lines use a 30$\times$30 (6$\times$6) $k$-mesh. \textbf{b)}~Normalized pSHG polar plots in parallel and perpendicular configurations under 1100 nm incident light. A shaded hexagon represents the estimated atomic orientation of both flakes. Individual plots for both polarizations are presented in Supplementary Figure 7.}
    \label{fig_pSHG}
\end{figure}

\section{Conclusion}

In summary, we have demonstrated effective inversion symmetry breaking in 2L TMD flakes through interlayer hybridization by artificial stacking.  The strong coupling leads to the formation of novel interlayer excitonic states and mechanical coupling, enabling sensitive probing via optical methods such as SHG, PL and Raman spectroscopy. The tetralayer heterostructure exhibited a robust SHG response comparable to that of 1L MoS$_2$, along with a distinct excitonic resonance landscape. By systematically varying the energy and polarization of the incident photons, we identified the evolution of the SHG signal strongly related to the IX contribution. Effective interlayer hybridization was also confirmed by theoretical estimations and the emergence of new phonon modes in Raman and excitonic features in PL measurements, presented by the LF and HF modes from lattice mismatch and the emergence of an interlayer excitonic PL peak.

Our observations highlight the effectiveness of artificial stacking as an approach to inversion symmetry breaking in van der Waals 2D  excitonic materials. The successful control of inversion symmetry offers a new tunable degree of freedom for tailoring the selection rules that govern the exotic physics of an innate centrosymmetric platform, significantly accelerating the development of applications in moiré photonics, integrated optoelectronics, and spin- and valleytronics. The potential enhanced tunability and control, together with the observed pSHG anisotropy, represent an advantage of tetralayer heterostructures as a robust nonlinear optical platform relative to stacked heterobilayers. More generally, our results are critical for developing strategies to design future miniaturized optoelectronic devices and excitonic technologies, making use of interlayer coupling to realize symmetry engineering and programmable light–matter interactions in 2D heterostructures.

\section{Experimental Methods}

\subsection{Sample fabrication} The heterostructure preparation involved artificially stacking mechanically exfoliated 2H-phase WSe$_2$ and MoS$_2$ flakes. The crystals were exfoliated using the scotch tape technique and thin flakes were produced using substrates of polydimethylsiloxane (PDMS). The stacking was carried out via PDMS-PDMS transfer of the obtained 2L flakes. This method is less optimal for thin flake obtention or heterostructure transfer, but successfully produces clean interfaces between flakes and ensures strong interlayer coupling. The fabricated heterostructure is then transferred on top of an Si substrate with 285~nm of thermally grown SiO$_2$. The Si substrate, with the heterostructure on top,  was placed on a hotplate at a temperature of 70~°C for 5 minutes during the PDMS-PDMS transfer and during the PDMS-Si transfer to maximize the uniformity of the process. The stacking angle between flakes was estimated from the alignment between the straight sides of the flakes, as the angle between intrinsic centrosymmetric flakes is not directly measurable by SHG.

\subsection{Linear optical characterizations} Raman and PL measurements were carried out with a 532~nm continuous wave laser. The Raman spectra were detected by a WITEC alpha 300 RA+ system equipped with a Newton Andor EMCCD, and a 100$\times$ CF Plan Nikon objective (NA =0.95). The visible PL signals were measured using a WITEC alpha 300 system, an iVac Andor OEM CCD camera and a 100$\times$ ZEISS Epiplan neofluar objective (NA =0.9). Lastly, the IR PL response was measured using our custom-built IR PL setup. A 532~nm laser beam from a laser diode (PicoQuant LDH-D-TA-530) was guided through a 50:50 beamsplitter into a 50$\times$ Olympus LCPlan N objective lens (0.65 NA) in reflection configuration. The reflected IR signal was isolated by a 750~nm longpass filter (Thorlabs FELH0750) and collected by an optical fiber (Thorlabs FG105LCA) into an Andor Kymera monochromator coupled to an Andor iDus camera.

\subsection{SHG characterization setup} SHG spectra were measured using our custom NLO setup. We implemented an ultrafast laser of 200 fs pulse width and 752 kHz repetition rate from a PHAROS/ORPHEUS (seed/OPA) system from Light Conversion. The light is directed to a reflective objective lens (40$\times$ Thorlabs LMM40XF-P01, 0.5 NA) through a 50:50 beamsplitter (Thorlabs BSW29R). The SHG signal is reflected after the beamsplitter and isolated from the incident wavelength with a 750~nm shortpass filter (Edmund \#64-332). Finally, the signal is collected with a multimode optical fiber (Thorlabs M92L02) into a Shamrock 500i spectrograph and detected by an Andor iXon Ultra EMCCD camera.  For all the single-point SHG measurements, the incident power at the sample was kept at 185 $\mu W$, with a laser spot size defined by the practical diffraction limit criterion for a Gaussian beam: $r=\frac{(0.67\lambda)}{NA}$. All measurements were normalized considering the quantum efficiency of the spectrometer and the detector, as well as the optical losses at each stage.

\subsection{SHG mapping setup} The complementary SHG mapping setup uses the same incident light as our NLO characterization setup. In this configuration, the light reaches our galvanometer scanner (Thorlabs GVS002) and passes through a pair of achromatic doublet lenses in telescope configuration. The light is then directed to a 50$\times$ Mitutoyo Plan Apochromat NIR (0.42 NA) objective lens through a 650~nm shortpass dichroic mirror (Thorlabs DMSP650R). The SHG signal is collected in reflection configuration by a second telescope after the dichroic mirror to reduce signal dispersion. Finally, the signal is cleared by a combination of a 1000~nm longpass dichroic mirror (Thorlabs DMLP1000R), a 550~nm longpass filter (Edmund \#62-984) and a 750~nm (Edmund \#64-332) shortpass filter, before being collected by a photomultiplier tube (Hamamatsu H7844). The signal was intentionally reduced to prevent saturation or damage in the detector. Incident power was kept at 185 $\mu W$ on the sample, with a laser spot size defined by the practical diffraction limit criterion for a Gaussian beam: $r=\frac{(0.67\lambda)}{NA}$.

\subsection{Computational methods}

The structures were relaxed with the Vienna \textit{ab initio} simulation package (VASP)~\cite{perdew1996generalized,kresse1996efficiency,kresse1996efficient} using the Perdew-Burke-Ernzerhof (PBE) functional with DFT-D3 van der Waals corrections, a plane-wave cutoff of 500~eV and a $\Gamma$-centred 12$\times$12$\times$1 $k$-mesh, with about 17.6~\AA\ of vacuum between periodic images. Atomic positions and the in-plane lattice constant were relaxed until the forces were below 0.02~eV~\AA$^{-1}$ (electronic convergence $10^{-6}$~eV). Band structures and second-order susceptibilities were calculated for the same structures with the all-electron full-potential LAPW code ELK (PBE with spin--orbit coupling). The independent-particle $\chi^{(2)}$ was computed in the length gauge~\cite{muller2020extending} on a 30$\times$30$\times$1 $k$-mesh with 240 empty states and a broadening of 0.082~eV. Excitonic effects were included by solving the Bethe--Salpeter equation in ELK on a 6$\times$6$\times$1 $k$-mesh (PBE+SOC, 16 valence and 16 conduction bands); from the resulting exciton states we computed $\chi^{(2)}$ with our own implementation of the excitonic second-order response (Supplementary Note 6). To give a physical explanation, we used a three-band tight-binding model~\cite{n5_liu_2013}, implemented with pyqula~\cite{n5_pyqula}, to compute the SHG with and without excitonic effects (Supplementary Note 5).

\section*{Supporting Information}

Supporting Information is available from the author.

\section*{Acknowledgments}

This work was supported by the Research Council of Finland (13357170, 360411, 359009, 365686, and 367808), the Research Council of Finland Flagship Programme (PREIN, Quantum), the Jane and Aatos Erkko foundation and the Technology Industries of Finland Centennial Foundation (Future Makers 2022) and the EU Marie Skłodowska-Curie Actions (ALTOS). This work is part of the Finnish Centre of Excellence in Quantum Materials (QMAT). H.F., and S.X. acknowledge the support from Villum Fonden (VIL70202, VIL78289).

\section*{Conflicts of Interest}

The authors declare no conflicts of interest.

\section*{Data Availability Statement}

Data supporting the findings of this study is available from the corresponding author upon reasonable request.

\bibliographystyle{unsrt}
\bibliography{Refs}

@article{zhao_origin_2013,
	title = {Origin of Indirect Optical Transitions in Few-Layer {MoS$_2$}, {WS$_2$}, and {WSe$_2$}},
	volume = {13},
	issn = {1530-6984},
	url = {https://doi.org/10.1021/nl403270k},
	doi = {10.1021/nl403270k},
	number = {11},
	urldate = {2025-11-25},
	journal = {Nano Letters},
	author = {Zhao, Weijie and Ribeiro, R. M. and Toh, Minglin and Carvalho, Alexandra and Kloc, Christian and Castro Neto, A. H. and Eda, Goki},
	month = nov,
	year = {2013},
	note = {Publisher: American Chemical Society},
	pages = {5627--5634},
}

@article{Tonndorf_photoluminescence_13,
author = {Philipp Tonndorf and Robert Schmidt and Philipp B\"{o}ttger and Xiao Zhang and Janna B\"{o}rner and Andreas Liebig and Manfred Albrecht and Christian Kloc and Ovidiu Gordan and Dietrich R. T. Zahn and Steffen Michaelis de Vasconcellos and Rudolf Bratschitsch},
journal = {Opt. Express},
number = {4},
pages = {4908--4916},
publisher = {Optica Publishing Group},
title = {Photoluminescence emission and {Raman} response of monolayer {MoS$_2$}, {MoSe$_2$}, and {WSe$_2$}},
volume = {21},
month = {Feb},
year = {2013},
url = {https://opg.optica.org/oe/abstract.cfm?URI=oe-21-4-4908},
doi = {10.1364/OE.21.004908},
}

@article{blaga_unveiling_2024,
	title = {Unveiling the complex phonon nature and phonon cascades in {1L} to {5L} {WSe}$_2$ using multiwavelength excitation {Raman} scattering},
	volume = {6},
	url = {https://pubs.rsc.org/en/content/articlelanding/2024/na/d4na00399c},
	doi = {10.1039/D4NA00399C},
	language = {en},
	number = {18},
	urldate = {2025-11-25},
	journal = {Nanoscale Advances},
	author = {Blaga, Claire and {\'A}lvarez, {\'A}ngel Labordet and Balgarkashi, Akshay and Banerjee, Mitali and {Fontcuberta i Morral}, Anna and Dimitrievska, Mirjana},
	year = {2024},
	note = {Publisher: Royal Society of Chemistry},
	pages = {4591--4603},
}

@article{obrien_mapping_2016,
	title = {Mapping of {Low}-{Frequency} {Raman} Modes in {CVD}-Grown Transition Metal Dichalcogenides: Layer Number, Stacking Orientation and Resonant Effects},
	volume = {6},
	copyright = {2016 The Author(s)},
	issn = {2045-2322},
	shorttitle = {Mapping of {Low}-{Frequency} {Raman} {Modes} in {CVD}-{Grown} {Transition} {Metal} {Dichalcogenides}},
	url = {https://www.nature.com/articles/srep19476},
	doi = {10.1038/srep19476},
	language = {en},
	number = {1},
	urldate = {2025-11-25},
	journal = {Scientific Reports},
	author = {O’Brien, Maria and McEvoy, Niall and Hanlon, Damien and Hallam, Toby and Coleman, Jonathan N. and Duesberg, Georg S.},
	month = jan,
	year = {2016},
	note = {Publisher: Nature Publishing Group},
	pages = {19476},
}

@article{zhao_atomically_2016,
	title = {Atomically phase-matched second-harmonic generation in a {2D} crystal},
	volume = {5},
	copyright = {2016 The Author(s)},
	issn = {2047-7538},
	url = {https://www.nature.com/articles/lsa2016131},
	doi = {10.1038/lsa.2016.131},
	language = {en},
	number = {8},
	urldate = {2025-11-25},
	journal = {Light: Science \& Applications},
	author = {Zhao, Mervin and Ye, Ziliang and Suzuki, Ryuji and Ye, Yu and Zhu, Hanyu and Xiao, Jun and Wang, Yuan and Iwasa, Yoshihiro and Zhang, Xiang},
	month = aug,
	year = {2016},
	note = {Publisher: Nature Publishing Group},
	pages = {e16131--e16131},
}

@article{shree_interlayer_2021,
	title = {Interlayer exciton mediated second harmonic generation in bilayer {MoS$_2$}},
	volume = {12},
	copyright = {2021 The Author(s)},
	issn = {2041-1723},
	url = {https://www.nature.com/articles/s41467-021-27213-8},
	doi = {10.1038/s41467-021-27213-8},
	language = {en},
	number = {1},
	urldate = {2025-11-25},
	journal = {Nature Communications},
	author = {Shree, Shivangi and Lagarde, Delphine and Lombez, Laurent and Robert, Cedric and Balocchi, Andrea and Watanabe, Kenji and Taniguchi, Takashi and Marie, Xavier and Gerber, Iann C. and Glazov, Mikhail M. and Golub, Leonid E. and Urbaszek, Bernhard and Paradisanos, Ioannis},
	month = nov,
	year = {2021},
	note = {Publisher: Nature Publishing Group},
	pages = {6894},
}

@article{perdew1996generalized,
  title={Generalized gradient approximation made simple},
  author={Perdew, John P and Burke, Kieron and Ernzerhof, Matthias},
  journal={Physical review letters},
  volume={77},
  number={18},
  pages={3865},
  year={1996},
  publisher={APS}
}

@article{kresse1996efficiency,
  title={Efficiency of ab-initio total energy calculations for metals and semiconductors using a plane-wave basis set},
  author={Kresse, Georg and Furthm{\"u}ller, J{\"u}rgen},
  journal={Computational materials science},
  volume={6},
  number={1},
  pages={15--50},
  year={1996},
  publisher={Elsevier}
}

@article{kresse1996efficient,
  title={Efficient iterative schemes for ab initio total-energy calculations using a plane-wave basis set},
  author={Kresse, Georg and Furthm{\"u}ller, J{\"u}rgen},
  journal={Physical review B},
  volume={54},
  number={16},
  pages={11169},
  year={1996},
  publisher={APS}
}

@article{muller2020extending,
  title={Extending solid-state calculations to ultra-long-range length scales},
  author={M{\"u}ller, Tristan and Sharma, Sangeeta and Gross, EKU and Dewhurst, JK},
  journal={Physical Review Letters},
  volume={125},
  number={25},
  pages={256402},
  year={2020},
  publisher={APS}
}

@article{shen_optical_1989,
	title = {Optical Second Harmonic Generation at Interfaces},
	volume = {40},
	language = {en},
	journal = {Annual Review Physical Chemistry},
	author = {Shen, Y R},
	month = oct,
	year = {1989},
	pages = {327--350},
}

@article{autere_optical_2018,
	title = {Optical harmonic generation in monolayer group-{VI} transition metal dichalcogenides},
	volume = {98},
	url = {https://link.aps.org/doi/10.1103/PhysRevB.98.115426},
	doi = {10.1103/PhysRevB.98.115426},
	number = {11},
	urldate = {2026-02-10},
	journal = {Physical Review B},
	author = {Autere, Anton and Jussila, Henri and Marini, Andrea and Saavedra, J. R. M. and Dai, Yunyun and Säynätjoki, Antti and Karvonen, Lasse and Yang, He and Amirsolaimani, Babak and Norwood, Robert A. and Peyghambarian, Nasser and Lipsanen, Harri and Kieu, Khanh and de Abajo, F. Javier García and Sun, Zhipei},
	month = sep,
	year = {2018},
	note = {Publisher: American Physical Society},
	pages = {115426},
}

@article{woodward_characterization_2016,
	title = {Characterization of the second- and third-order nonlinear optical susceptibilities of monolayer {MoS$_2$} using multiphoton microscopy},
	volume = {4},
	issn = {2053-1583},
	url = {https://iopscience.iop.org/article/10.1088/2053-1583/4/1/011006},
	doi = {10.1088/2053-1583/4/1/011006},
	language = {en},
	number = {1},
	urldate = {2026-02-10},
	journal = {2D Materials},
	author = {Woodward, R I and Murray, R T and Phelan, C F and Oliveira, R E P De and Runcorn, T H and Kelleher, E J R and Li, S and Oliveira, E C De and Fechine, G J M and Eda, G and Matos, C J S De},
	month = nov,
	year = {2016},
	pages = {011006},
}

@article{yan_stacking-dependent_2015,
	title = {Stacking-Dependent Interlayer Coupling in Trilayer {MoS$_2$} with Broken Inversion Symmetry},
	volume = {15},
	issn = {1530-6984},
	url = {https://doi.org/10.1021/acs.nanolett.5b03597},
	doi = {10.1021/acs.nanolett.5b03597},
	number = {12},
	urldate = {2026-03-24},
	journal = {Nano Letters},
	author = {Yan, Jiaxu and Xia, Juan and Wang, Xingli and Liu, Lei and Kuo, Jer-Lai and Tay, Beng Kang and Chen, Shoushun and Zhou, Wu and Liu, Zheng and Shen, Ze Xiang},
	month = dec,
	year = {2015},
	note = {Publisher: American Chemical Society},
	pages = {8155--8161},
}

@article{li_probing_2013,
	title = {Probing Symmetry Properties of Few-Layer {MoS$_2$} and h-{BN} by Optical Second-Harmonic Generation},
	volume = {13},
	issn = {1530-6984},
	url = {https://doi.org/10.1021/nl401561r},
	doi = {10.1021/nl401561r},
	number = {7},
	urldate = {2026-03-30},
	journal = {Nano Letters},
	author = {Li, Yilei and Rao, Yi and Mak, Kin Fai and You, Yumeng and Wang, Shuyuan and Dean, Cory R. and Heinz, Tony F.},
	month = jul,
	year = {2013},
	note = {Publisher: American Chemical Society},
	pages = {3329--3333},
}

@article{chiu_determination_2015,
	title = {Determination of band alignment in the single-layer {MoS$_2$}/{WSe$_2$} heterojunction},
	volume = {6},
	copyright = {2015 The Author(s)},
	issn = {2041-1723},
	url = {https://www.nature.com/articles/ncomms8666},
	doi = {10.1038/ncomms8666},
	language = {en},
	number = {1},
	urldate = {2026-03-31},
	journal = {Nature Communications},
	author = {Chiu, Ming-Hui and Zhang, Chendong and Shiu, Hung-Wei and Chuu, Chih-Piao and Chen, Chang-Hsiao and Chang, Chih-Yuan S. and Chen, Chia-Hao and Chou, Mei-Yin and Shih, Chih-Kang and Li, Lain-Jong},
	month = jul,
	year = {2015},
	note = {Publisher: Nature Publishing Group},
	pages = {7666},
}

@article{aslan_strain_2018,
	title = {Strain tuning of excitons in monolayer {WSe$_2$}},
	volume = {98},
	url = {https://link.aps.org/doi/10.1103/PhysRevB.98.115308},
	doi = {10.1103/PhysRevB.98.115308},
	number = {11},
	urldate = {2026-04-01},
	journal = {Physical Review B},
	author = {Aslan, Burak and Deng, Minda and Heinz, Tony F.},
	month = sep,
	year = {2018},
	note = {Publisher: American Physical Society},
	pages = {115308},
}

@article{yang_tunable_2020,
	title = {Tunable Second Harmonic Generation in Twisted Bilayer Graphene},
	volume = {3},
	issn = {2590-2385},
	url = {https://www.sciencedirect.com/science/article/pii/S2590238520304458},
	doi = {10.1016/j.matt.2020.08.018},
	number = {4},
	urldate = {2026-04-08},
	journal = {Matter},
	author = {Yang, Fuyi and Song, Wenshen and Meng, Fanhao and Luo, Fuchuan and Lou, Shuai and Lin, Shuren and Gong, Zilun and Cao, Jinhua and Barnard, Edward S. and Chan, Emory and Yang, Li and Yao, Jie},
	month = oct,
	year = {2020},
	pages = {1361--1376},
}

@article{cao_unconventional_2018,
	title = {Unconventional superconductivity in magic-angle graphene superlattices},
	volume = {556},
	copyright = {2018 Macmillan Publishers Limited, part of Springer Nature. All rights reserved.},
	issn = {1476-4687},
	url = {https://www.nature.com/articles/nature26160},
	doi = {10.1038/nature26160},
	language = {en},
	number = {7699},
	urldate = {2026-04-08},
	journal = {Nature},
	author = {Cao, Yuan and Fatemi, Valla and Fang, Shiang and Watanabe, Kenji and Taniguchi, Takashi and Kaxiras, Efthimios and Jarillo-Herrero, Pablo},
	month = apr,
	year = {2018},
	note = {Publisher: Nature Publishing Group},
	pages = {43--50},
}

@article{lin_moire-engineered_2024,
	title = {Moiré-engineered light-matter interactions in {MoS$_2$}/{WSe$_2$} heterobilayers at room temperature},
	volume = {15},
	copyright = {2024 The Author(s)},
	issn = {2041-1723},
	url = {https://www.nature.com/articles/s41467-024-53083-x},
	doi = {10.1038/s41467-024-53083-x},
	language = {en},
	number = {1},
	urldate = {2026-04-09},
	journal = {Nature Communications},
	author = {Lin, Qiaoling and Fang, Hanlin and Kalaboukhov, Alexei and Liu, Yuanda and Zhang, Yi and Fischer, Moritz and Li, Juntao and Hagel, Joakim and Brem, Samuel and Malic, Ermin and Stenger, Nicolas and Sun, Zhipei and Wubs, Martijn and Xiao, Sanshui},
	month = oct,
	year = {2024},
	note = {Publisher: Nature Publishing Group},
	pages = {8762},
}

@article{tan_layer-engineered_2021,
	title = {Layer-engineered interlayer excitons},
	volume = {7},
	url = {https://www.science.org/doi/10.1126/sciadv.abh0863},
	doi = {10.1126/sciadv.abh0863},
	number = {30},
	urldate = {2026-04-09},
	journal = {Science Advances},
	author = {Tan, Qinghai and Rasmita, Abdullah and Li, Si and Liu, Sheng and Huang, Zumeng and Xiong, Qihua and Yang, Shengyuan A. and Novoselov, K. S. and Gao, Wei-bo},
	month = jul,
	year = {2021},
	note = {Publisher: American Association for the Advancement of Science},
	pages = {eabh0863},
}

@article{zhu_observation_2015,
	title = {Observation of piezoelectricity in free-standing monolayer {MoS$_2$}},
	volume = {10},
	issn = {1748-3395},
	url = {https://www.nature.com/articles/nnano.2014.309},
	doi = {10.1038/nnano.2014.309},
	number = {2},
	journal = {Nature Nanotechnology},
	author = {Zhu, Hanyu and Wang, Yuan and Xiao, Jun and Liu, Ming and Xiong, Shaomin and Wong, Zi Jing and Ye, Ziliang and Ye, Yu and Yin, Xiaobo and Zhang, Xiang},
	month = feb,
	year = {2015},
	pages = {151--155},
}

@article{wu_piezoelectricity_2014,
	title = {Piezoelectricity of single-atomic-layer {MoS$_2$} for energy conversion and piezotronics},
	volume = {514},
	issn = {1476-4687},
	url = {https://www.nature.com/articles/nature13792},
	doi = {10.1038/nature13792},
	number = {7523},
	journal = {Nature},
	author = {Wu, Wenzhuo and Wang, Lei and Li, Yilei and Zhang, Fan and Lin, Long and Niu, Simiao and Chenet, Daniel and Zhang, Xian and Hao, Yufeng and Heinz, Tony F. and Hone, James and Wang, Zhong Lin},
	month = oct,
	year = {2014},
	pages = {470--474},
}

@article{yasuda_stacking_2021,
	title = {Stacking-engineered ferroelectricity in bilayer boron nitride},
	volume = {372},
	issn = {0036-8075},
	url = {https://www.science.org/doi/10.1126/science.abd3230},
	doi = {10.1126/science.abd3230},
	number = {6549},
	journal = {Science},
	author = {Yasuda, Kenji and Wang, Xirui and Watanabe, Kenji and Taniguchi, Takashi and Jarillo-Herrero, Pablo},
	month = jun,
	year = {2021},
	pages = {1458--1462},
}

@article{vizner_interfacial_2021,
	title = {Interfacial ferroelectricity by van der {Waals} sliding},
	volume = {372},
	issn = {0036-8075},
	url = {https://www.science.org/doi/10.1126/science.abe8177},
	doi = {10.1126/science.abe8177},
	number = {6549},
	journal = {Science},
	author = {Vizner Stern, Moshe and Waschitz, Yuval and Cao, Wei and Nevo, Ido and Watanabe, Kenji and Taniguchi, Takashi and Sela, Erez and Urbakh, Michael and Hod, Oded and Ben Shalom, Moshe},
	month = jun,
	year = {2021},
	pages = {1462--1466},
}

@article{fei_ferroelectric_2018,
	title = {Ferroelectric switching of a two-dimensional metal},
	volume = {560},
	issn = {1476-4687},
	url = {https://www.nature.com/articles/s41586-018-0336-3},
	doi = {10.1038/s41586-018-0336-3},
	number = {7718},
	journal = {Nature},
	author = {Fei, Zaiyao and Zhao, Wenjin and Palomaki, Tauno A. and Sun, Bosong and Miller, Moira K. and Zhao, Zhiying and Yan, Jiaqiang and Xu, Xiaodong and Cobden, David H.},
	month = aug,
	year = {2018},
	pages = {336--339},
}

@article{mak_valley_2012,
	title = {Control of valley polarization in monolayer {MoS$_2$} by optical helicity},
	volume = {7},
	issn = {1748-3387},
	url = {https://www.nature.com/articles/nnano.2012.96},
	doi = {10.1038/nnano.2012.96},
	number = {8},
	journal = {Nature Nanotechnology},
	author = {Mak, Kin Fai and He, Keliang and Shan, Jie and Heinz, Tony F.},
	month = aug,
	year = {2012},
	pages = {494--498},
}

@article{cao_valley-selective_2012,
	title = {Valley-selective circular dichroism of monolayer molybdenum disulphide},
	volume = {3},
	issn = {2041-1723},
	url = {https://www.nature.com/articles/ncomms1882},
	doi = {10.1038/ncomms1882},
	journal = {Nature Communications},
	author = {Cao, Ting and Wang, Gang and Han, Wenpeng and Ye, Huiqi and Zhu, Chuanrui and Shi, Junren and Niu, Qian and Tan, Peihong and Wang, Enge and Liu, Baoli and Feng, Ji},
	month = may,
	year = {2012},
	pages = {887},
}

@article{novoselov_2d_2016,
	title = {{2D} materials and van der {Waals} heterostructures},
	volume = {353},
	issn = {0036-8075},
	url = {https://www.science.org/doi/10.1126/science.aac9439},
	doi = {10.1126/science.aac9439},
	number = {6298},
	journal = {Science},
	author = {Novoselov, K. S. and Mishchenko, A. and Carvalho, A. and Castro Neto, A. H.},
	month = jul,
	year = {2016},
	pages = {aac9439},
}

@article{paradisanos_second_2022,
	title = {Second harmonic generation control in twisted bilayers of transition metal dichalcogenides},
	volume = {105},
	issn = {2469-9950, 2469-9969},
	url = {https://link.aps.org/doi/10.1103/PhysRevB.105.115420},
	doi = {10.1103/PhysRevB.105.115420},
	language = {en},
	number = {11},
	urldate = {2026-04-15},
	journal = {Physical Review B},
	author = {Paradisanos, Ioannis and Raven, Andres Manuel Saiz and Amand, Thierry and Robert, Cedric and Renucci, Pierre and Watanabe, Kenji and Taniguchi, Takashi and Gerber, Iann C. and Marie, Xavier and Urbaszek, Bernhard},
	month = mar,
	year = {2022},
	pages = {115420},
}

@article{arias-munoz_interlayer_2024,
	title = {Interlayer Coupling Limit in Artificially Stacked {MoS$_2$} Homojunctions},
	volume = {34},
	copyright = {© 2023 The Authors. Advanced Functional Materials published by Wiley-VCH GmbH},
	issn = {1616-3028},
	url = {https://onlinelibrary.wiley.com/doi/abs/10.1002/adfm.202310365},
	doi = {10.1002/adfm.202310365},
	language = {en},
	number = {11},
	urldate = {2025-11-25},
	journal = {Advanced Functional Materials},
	author = {Arias-Muñoz, Juan C. and Kaaripuro, Henri and Zhang, Yi and Das, Susobhan and Fernandez, Henry A. and Sun, Zhipei},
	year = {2024},
	note = {\_eprint: https://advanced.onlinelibrary.wiley.com/doi/pdf/10.1002/adfm.202310365},
	pages = {2310365},
}

@article{du_engineering_2021,
	title = {Engineering symmetry breaking in {2D} layered materials},
	volume = {3},
	copyright = {2021 Springer Nature Limited},
	issn = {2522-5820},
	url = {https://www.nature.com/articles/s42254-020-00276-0},
	doi = {10.1038/s42254-020-00276-0},
	language = {en},
	number = {3},
	urldate = {2025-12-09},
	journal = {Nature Reviews Physics},
	author = {Du, Luojun and Hasan, Tawfique and Castellanos-Gomez, Andres and Liu, Gui-Bin and Yao, Yugui and Lau, Chun Ning and Sun, Zhipei},
	month = mar,
	year = {2021},
	note = {Publisher: Nature Publishing Group},
	pages = {193--206},
}

@article{autere_nonlinear_2018,
	title = {Nonlinear Optics with {2D} Layered Materials},
	volume = {30},
	copyright = {© 2018 WILEY-VCH Verlag GmbH \& Co. KGaA, Weinheim},
	issn = {1521-4095},
	url = {https://onlinelibrary.wiley.com/doi/abs/10.1002/adma.201705963},
	doi = {10.1002/adma.201705963},
	language = {en},
	number = {24},
	urldate = {2025-11-20},
	journal = {Advanced Materials},
	author = {Autere, Anton and Jussila, Henri and Dai, Yunyun and Wang, Yadong and Lipsanen, Harri and Sun, Zhipei},
	year = {2018},
	note = {\_eprint: https://advanced.onlinelibrary.wiley.com/doi/pdf/10.1002/adma.201705963},
	pages = {1705963},
}

@article{hsu_second_2014,
	title = {Second Harmonic Generation from Artificially Stacked Transition Metal Dichalcogenide Twisted Bilayers},
	volume = {8},
	issn = {1936-0851},
	url = {https://doi.org/10.1021/nn500228r},
	doi = {10.1021/nn500228r},
	number = {3},
	urldate = {2025-11-20},
	journal = {ACS Nano},
	author = {Hsu, Wei-Ting and Zhao, Zi-Ang and Li, Lain-Jong and Chen, Chang-Hsiao and Chiu, Ming-Hui and Chang, Pi-Shan and Chou, Yi-Chia and Chang, Wen-Hao},
	month = mar,
	year = {2014},
	note = {Publisher: American Chemical Society},
	pages = {2951--2958},
}

@article{zhang_coherent_2022,
	title = {Coherent modulation of chiral nonlinear optics with crystal symmetry},
	volume = {11},
	copyright = {2022 The Author(s)},
	issn = {2047-7538},
	url = {https://www.nature.com/articles/s41377-022-00915-4},
	doi = {10.1038/s41377-022-00915-4},
	language = {en},
	number = {1},
	urldate = {2025-11-20},
	journal = {Light: Science \& Applications},
	author = {Zhang, Yi and Bai, Xueyin and Arias Muñoz, Juan and Dai, Yunyun and Das, Susobhan and Wang, Yadong and Sun, Zhipei},
	month = jul,
	year = {2022},
	note = {Publisher: Nature Publishing Group},
	pages = {216},
}

@article{turunen_quantum_2022,
	title = {Quantum photonics with layered {2D} materials},
	volume = {4},
	copyright = {2022 Springer Nature Limited},
	issn = {2522-5820},
	url = {https://www.nature.com/articles/s42254-021-00408-0},
	doi = {10.1038/s42254-021-00408-0},
	language = {en},
	number = {4},
	urldate = {2025-11-19},
	journal = {Nature Reviews Physics},
	author = {Turunen, Mikko and Brotons-Gisbert, Mauro and Dai, Yunyun and Wang, Yadong and Scerri, Eleanor and Bonato, Cristian and Jöns, Klaus D. and Sun, Zhipei and Gerardot, Brian D.},
	month = apr,
	year = {2022},
	note = {Publisher: Nature Publishing Group},
	pages = {219--236},
}

@article{kim_anomalous_2022,
	title = {Anomalous optical excitations from arrays of whirlpooled lattice distortions in moiré superlattices},
	volume = {21},
	copyright = {2022 The Author(s), under exclusive licence to Springer Nature Limited},
	issn = {1476-4660},
	url = {https://www.nature.com/articles/s41563-022-01240-2},
	doi = {10.1038/s41563-022-01240-2},
	language = {en},
	number = {8},
	urldate = {2025-12-01},
	journal = {Nature Materials},
	author = {Kim, Jungcheol and Ko, Eunjung and Jo, Jaeyeon and Kim, Miyoung and Yoo, Hyobin and Son, Young-Woo and Cheong, Hyeonsik},
	month = aug,
	year = {2022},
	note = {Publisher: Nature Publishing Group},
	pages = {890--895},
}

@article{yagmurcukardes_strain_2018,
	title = {Strain mapping in single-layer two-dimensional crystals via {Raman} activity},
	volume = {97},
	url = {https://link.aps.org/doi/10.1103/PhysRevB.97.115427},
	doi = {10.1103/PhysRevB.97.115427},
	number = {11},
	urldate = {2025-12-01},
	journal = {Physical Review B},
	author = {Yagmurcukardes, M. and Bacaksiz, C. and Unsal, E. and Akbali, B. and Senger, R. T. and Sahin, H.},
	month = mar,
	year = {2018},
	note = {Publisher: American Physical Society},
	pages = {115427},
}

@article{velicky_strain_2020,
	title = {Strain and Charge Doping Fingerprints of the Strong Interaction between Monolayer {MoS$_2$} and Gold},
	volume = {11},
	url = {https://doi.org/10.1021/acs.jpclett.0c01287},
	doi = {10.1021/acs.jpclett.0c01287},
	number = {15},
	urldate = {2025-12-01},
	journal = {The Journal of Physical Chemistry Letters},
	author = {Velický, Matěj and Rodriguez, Alvaro and Bouša, Milan and Krayev, Andrey V. and Vondráček, Martin and Honolka, Jan and Ahmadi, Mahdi and Donnelly, Gavin E. and Huang, Fumin and Abruña, Héctor D. and Novoselov, Kostya S. and Frank, Otakar},
	month = aug,
	year = {2020},
	note = {Publisher: American Chemical Society},
	pages = {6112--6118},
}

@article{doan_charge_2017,
	title = {Charge Transport in {MoS$_2$}/{WSe$_2$} van der {Waals} Heterostructure with Tunable Inversion Layer},
	volume = {11},
	issn = {1936-0851, 1936-086X},
	url = {https://pubs.acs.org/doi/10.1021/acsnano.7b00021},
	doi = {10.1021/acsnano.7b00021},
	language = {en},
	number = {4},
	urldate = {2025-12-01},
	journal = {ACS Nano},
	author = {Doan, Manh-Ha and Jin, Youngjo and Adhikari, Subash and Lee, Sanghyub and Zhao, Jiong and Lim, Seong Chu and Lee, Young Hee},
	month = apr,
	year = {2017},
	pages = {3832--3840},
}

@article{li_efficient_2020,
	title = {Efficient strain modulation of {2D} materials via polymer encapsulation},
	volume = {11},
	copyright = {2020 The Author(s)},
	issn = {2041-1723},
	url = {https://www.nature.com/articles/s41467-020-15023-3},
	doi = {10.1038/s41467-020-15023-3},
	language = {en},
	number = {1},
	urldate = {2025-12-02},
	journal = {Nature Communications},
	author = {Li, Zhiwei and Lv, Yawei and Ren, Liwang and Li, Jia and Kong, Lingan and Zeng, Yujia and Tao, Quanyang and Wu, Ruixia and Ma, Huifang and Zhao, Bei and Wang, Di and Dang, Weiqi and Chen, Keqiu and Liao, Lei and Duan, Xidong and Duan, Xiangfeng and Liu, Yuan},
	month = mar,
	year = {2020},
	note = {Publisher: Nature Publishing Group},
	pages = {1151},
}

@article{zhang_emergent_2023,
	title = {Emergent second-harmonic generation in van der {Waals} heterostructure of bilayer {MoS$_2$} and monolayer graphene},
	volume = {9},
	url = {https://www.science.org/doi/10.1126/sciadv.adf4571},
	doi = {10.1126/sciadv.adf4571},
	number = {11},
	urldate = {2025-11-25},
	journal = {Science Advances},
	author = {Zhang, Mingwen and Han, Nannan and Zhang, Jiachen and Wang, Jing and Chen, Xiaoqing and Zhao, Jianlin and Gan, Xuetao},
	month = mar,
	year = {2023},
	note = {Publisher: American Association for the Advancement of Science},
	pages = {eadf4571},
}

@article{kosmider_large_2013,
	title = {Large spin splitting in the conduction band of transition metal dichalcogenide monolayers},
	volume = {88},
	url = {https://link.aps.org/doi/10.1103/PhysRevB.88.245436},
	doi = {10.1103/PhysRevB.88.245436},
	number = {24},
	urldate = {2025-11-25},
	journal = {Physical Review B},
	author = {Kośmider, K. and González, J. W. and Fernández-Rossier, J.},
	month = dec,
	year = {2013},
	note = {Publisher: American Physical Society},
	pages = {245436},
}

@article{arora_excitonic_2015,
	title = {Excitonic resonances in thin films of {WSe$_2$}: from monolayer to bulk material},
	volume = {7},
	issn = {2040-3372},
	shorttitle = {Excitonic resonances in thin films of {WSe2}},
	url = {https://pubs.rsc.org/en/content/articlelanding/2015/nr/c5nr01536g},
	doi = {10.1039/C5NR01536G},
	language = {en},
	number = {23},
	urldate = {2025-11-25},
	journal = {Nanoscale},
	author = {Arora, Ashish and Koperski, Maciej and Nogajewski, Karol and Marcus, Jacques and Faugeras, Clément and Potemski, Marek},
	month = jun,
	year = {2015},
	note = {Publisher: The Royal Society of Chemistry},
	pages = {10421--10429},
}

@article{kunstmann_momentum-space_2018,
	title = {Momentum-space indirect interlayer excitons in transition-metal dichalcogenide van der {Waals} heterostructures},
	volume = {14},
	copyright = {2018 The Author(s)},
	issn = {1745-2481},
	url = {https://www.nature.com/articles/s41567-018-0123-y},
	doi = {10.1038/s41567-018-0123-y},
	language = {en},
	number = {8},
	urldate = {2025-11-25},
	journal = {Nature Physics},
	author = {Kunstmann, Jens and Mooshammer, Fabian and Nagler, Philipp and Chaves, Andrey and Stein, Frederick and Paradiso, Nicola and Plechinger, Gerd and Strunk, Christoph and Schüller, Christian and Seifert, Gotthard and Reichman, David R. and Korn, Tobias},
	month = aug,
	year = {2018},
	note = {Publisher: Nature Publishing Group},
	pages = {801--805},
}

@article{splendiani_emerging_2010,
	title = {Emerging Photoluminescence in Monolayer {MoS$_2$}},
	volume = {10},
	issn = {1530-6984, 1530-6992},
	url = {https://pubs.acs.org/doi/10.1021/nl903868w},
	doi = {10.1021/nl903868w},
	language = {en},
	number = {4},
	urldate = {2025-11-25},
	journal = {Nano Letters},
	author = {Splendiani, Andrea and Sun, Liang and Zhang, Yuanbo and Li, Tianshu and Kim, Jonghwan and Chim, Chi-Yung and Galli, Giulia and Wang, Feng},
	month = apr,
	year = {2010},
	pages = {1271--1275},
}

@article{sun_optical_2016,
	title = {Optical modulators with {2D} layered materials},
	volume = {10},
	copyright = {2016 Springer Nature Limited},
	issn = {1749-4893},
	url = {https://www.nature.com/articles/nphoton.2016.15},
	doi = {10.1038/nphoton.2016.15},
	language = {en},
	number = {4},
	urldate = {2026-04-16},
	journal = {Nature Photonics},
	author = {Sun, Zhipei and Martinez, Amos and Wang, Feng},
	month = apr,
	year = {2016},
	note = {Publisher: Nature Publishing Group},
	pages = {227--238},
}

@article{chen_luminescent_2019,
	title = {Luminescent Emission of Excited Rydberg Excitons from Monolayer {WSe$_2$}},
	volume = {19},
	issn = {1530-6984},
	url = {https://doi.org/10.1021/acs.nanolett.9b00029},
	doi = {10.1021/acs.nanolett.9b00029},
	number = {4},
	urldate = {2026-04-20},
	journal = {Nano Letters},
	author = {Chen, Shao-Yu and Lu, Zhengguang and Goldstein, Thomas and Tong, Jiayue and Chaves, Andrey and Kunstmann, Jens and Cavalcante, L. S. R. and Woźniak, Tomasz and Seifert, Gotthard and Reichman, D. R. and Taniguchi, Takashi and Watanabe, Kenji and Smirnov, Dmitry and Yan, Jun},
	month = apr,
	year = {2019},
	note = {Publisher: American Chemical Society},
	pages = {2464--2471},
}

@misc{tang_forbidden_2026,
	title = {Forbidden second harmonics in centrosymmetric bilayer crystals},
	copyright = {arXiv.org perpetual, non-exclusive license},
	url = {https://arxiv.org/abs/2601.08830},
	doi = {10.48550/ARXIV.2601.08830},
	language = {en},
	urldate = {2026-04-22},
	publisher = {arXiv},
	author = {Tang, Haoning and Ding, Zhitong and Ruan, Tianyi and Hao, Zeyu and Watanabe, Kenji and Taniguchi, Takashi and Wang, Haozhe and Javey, Ali and Wang, Feng and Cao, Yuan},
	year = {2026},
	note = {Version Number: 1},
}

@article{wang_differential_2026,
	title = {Differential modulation effect of rare earth ions on interfacial synergy and electromagnetic absorption of {RE}$_{\textrm{2}}${O}$_{\textrm{3}}$@{MXene} heterostructures},
	volume = {14},
	issn = {2050-7488, 2050-7496},
	url = {http://pubs.rsc.org/TA/article-lookup/doi/10.1039/D6TA05699G},
	doi = {10.1039/D6TA05699G},
	language = {en},
	number = {49},
	urldate = {2026-10-10},
	journal = {Journal of Materials Chemistry A},
	author = {Wang, Qiang and Su, Xiaolei and Liu, Yi},
	year = {2026},
	pages = {33336--33353},
}

@misc{n5_pyqula,
  author       = {Lado, Jose L.},
  title        = {pyqula: a {P}ython library for quantum lattice models},
  howpublished = {\url{https://github.com/joselado/pyqula}},
  year         = {2026}
}

@article{n5_liu_2013,
  author  = {Liu, Gui-Bin and Shan, Wen-Yu and Yao, Yugui and Yao, Wang and Xiao, Di},
  title   = {Three-band tight-binding model for monolayers of group-{VIB} transition metal dichalcogenides},
  journal = {Phys. Rev. B},
  volume  = {88},
  pages   = {085433},
  year    = {2013},
  note    = {Erratum: Phys. Rev. B 89, 039901 (2014)},
  doi     = {10.1103/PhysRevB.88.085433}
}

\end{document}